\documentclass[a4paper,fleqn]{cas-dc}

\usepackage[authoryear,longnamesfirst]{natbib}
\usepackage{amssymb}
\usepackage{amsmath}
\usepackage{amsfonts}
\usepackage{graphicx}
\usepackage{nicefrac}
\usepackage{float}
\usepackage{hyperref}
\usepackage{epigraph}
\usepackage{csquotes}
\usepackage[utf8]{inputenc}
\usepackage[english]{babel}
\usepackage{amsthm,thmtools}
\usepackage[nottoc]{tocbibind}
\usepackage{caption}
\usepackage{subcaption}
\usepackage{longtable}
\usepackage{booktabs}
\usepackage[table,xcdraw]{xcolor}
\usepackage{pdflscape}
\usepackage{rotating}
\usepackage{multirow}
\usepackage{physics}
\usepackage[mathscr]{euscript}

\def\tsc#1{\csdef{#1}{\textsc{\lowercase{#1}}\xspace}}
\tsc{WGM}
\tsc{QE}
\begin{document}
\let\WriteBookmarks\relax
\def\floatpagepagefraction{1}
\def\textpagefraction{.001}

\shorttitle{}    

\shortauthors{}  

\title [mode = title]{A Methodological Study on Data Representation for Machine Learning Modelling of Thermal Conductivity of Rare-Earth Oxides}  

\tnotemark[1] 

\tnotetext[1]{} 

\author[1]{Amiya Chowdhury}

\cormark[1]

\fnmark[1]

\ead{amiya.chowdhury@nottingham.ac.uk}

\ead[url]{https://orcid.org/0009-0004-2537-4962}

\credit{Conceptualisation, Methodology, Software, Formal Analysis, Data curation, Investigation, Writing - Original Draft Preparation}

\affiliation[1]{organization={Centre for Excellence in Coatings and Surface Engineering, University of Nottingham},
            addressline={University Park}, 
            city={Nottingham},
            postcode={NG7 2RD}, 
            country={UK}}

\author[2]{Acacio Rinc\'on Romero}

% Footnote of the second author
\fnmark[2]

% Email id of the second author
\ead{}

\ead[url]{}

\credit{Conceptualisation, Methodology, Data curation, Supervision, Writing - Reviewing and Editing}

\affiliation[1]{organization={Centre for Excellence in Coatings and Surface Engineering, University of Nottingham},
            addressline={University Park}, 
            city={Nottingham},
            postcode={NG7 2RD}, 
            country={UK}}

\author[3]{Eduardo Aguilar-Bejarano}

% Footnote of the second author
\fnmark[3]

% Email id of the second author
\ead{}

\ead[url]{}

\credit{Methodology, Software}

\affiliation[2]{organization={School of Chemistry, University of Nottingham},
            addressline={Jubilee Campus}, 
            city={Nottingham},
%          citysep={}, % Uncomment if no comma needed between city and postcode
            postcode={NG8 1BB}, 
            state={},
            country={UK}}

\author[4]{Halar Memon}

% Footnote of the second author
\fnmark[4]

% Email id of the second author
\ead{}

\ead[url]{}

\credit{Writing - Reviewing and Editing}

\affiliation[1]{organization={Centre for Excellence in Coatings and Surface Engineering, University of Nottingham},
            addressline={University Park}, 
            city={Nottingham},
            postcode={NG7 2RD}, 
            country={UK}}

\author[5]{Grazziela Figueredo}

% Footnote of the second author
\fnmark[5]

% Email id of the second author
\ead{}

\ead[url]{G.Figueredo@nottingham.ac.uk}

\credit{Conceptualisation, Methodology, Software, Supervision, Writing - Reviewing and Editing}

\affiliation[3]{organization={Centre for Health Informatics, University of Nottingham},
            addressline={University Park}, 
            city={Nottingham},
            postcode={NG7 2RD}, 
            country={UK}}

\author[6]{Tanvir Hussain}

% Footnote of the second author
\fnmark[6]

% Email id of the second author
\ead{tanvir.hussain@nottingham.ac.uk}

\ead[url]{}

\credit{Conceptualisation, Fund acquisition, Supervision, Writing - Reviewing and Editing}

\affiliation[1]{organization={Centre for Excellence in Coatings and Surface Engineering, University of Nottingham},
            addressline={University Park}, 
            city={Nottingham},
            postcode={NG7 2RD}, 
            country={UK}}

\cortext[1]{Corresponding author}

\fntext[1]{}


\begin{abstract}
Quantitative structure-activity relationship (QSAR) modelling is widely employed in materials science to predict properties of interest and extract useful descriptors for measured properties. In thermal barrier coatings (TBC), QSAR can significantly shorten the experimental discovery cycle, which can take years. Although machine learning methods are commonly employed for QSAR, their performance depends on the data quality and how instances are represented. Traditional, hand-crafted descriptors based on known material properties are limited to represent materials that share the same basic crystal structure, limited the size of the dataset. By contrast, graph neural networks offer a more expressive representation, encoding atomic positions and bonds in the crystal lattice. In this study, we compare Random Forest (RF) and Gaussian Process (GP) models trained on hand-crafted descriptors from the literature with graph-based representations for high-entropy, rare-earth pyrochlore oxides using the Crystal Graph Convolutional Neural Network (CGCNN). Two different types of augmentation methods are also explored to account for the limited data size, one of which is only applicable to graph-based representations. Our findings show that the CGCNN model substantially outperforms the RF and GP models, underscoring the potential of graph-based representations for enhanced QSAR modelling in TBC research.
\end{abstract}

\begin{graphicalabstract}
\includegraphics[width=1\textwidth]{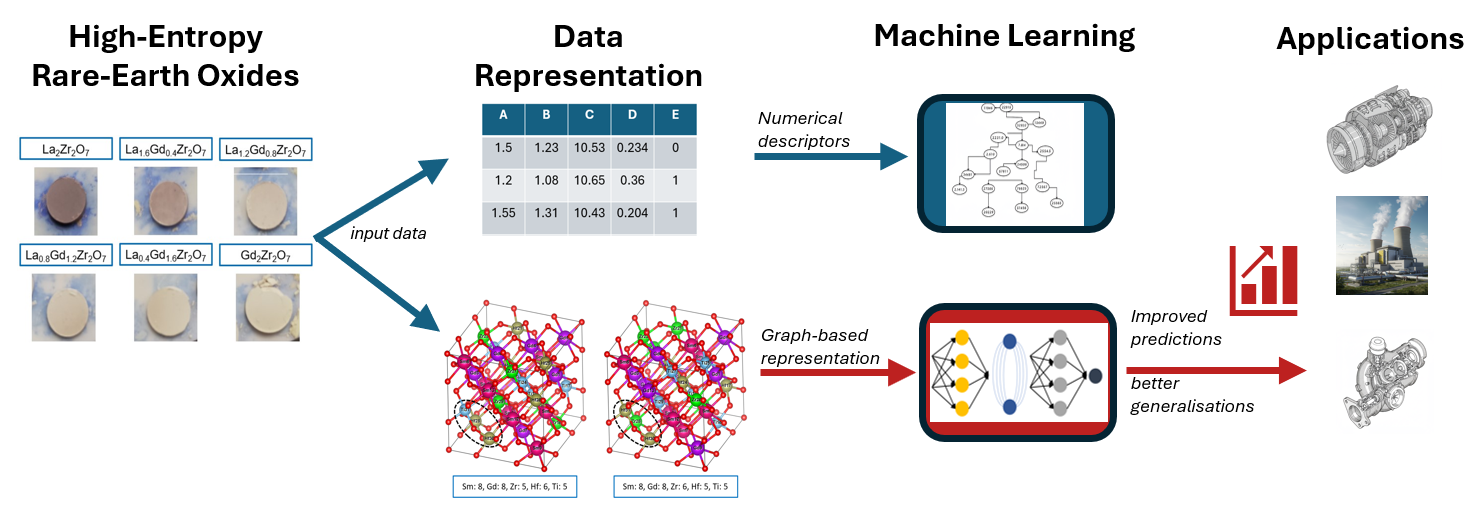}
\end{graphicalabstract}

\begin{highlights}
\item Use of graph neural networks to model and predict thermal conductivity for high-entropy rare-earth oxides
\item Comparison of graph data representation of compositions with handcrafted descriptors found in the literature
\item Experiments show improved performance and generalisation with graph representations, demonstrating its suitability for modelling the problem
\end{highlights}

%\nocite{*}

% Keywords

\maketitle

% Main text
\section{Introduction}\label{Intro}

Operating temperatures for nickel based superalloys are between 800-1000$^oC$ (\cite{superalloys_Mouritz_2012b}). Turbine inlet temperatures in gas-turbine engines can exceed 1500$^oC$ (\cite{Gas_turbine_engine_Centrich_2014}).Thermal barrier coatings (TBC) provide temperature reductions on the nickel-based superalloy of 100-200$^oC$ (\cite{darolia_2013}), allowing turbine engines to operate effectively at higher temperatures. Higher engine operating temperatures lead to increased thermodynamic efficiency and lower $CO_2$ emissions. Yttria stabilised zirconia has been the industry standard TBC since the 1970s due to its combination of thermal and mechanical properties (\cite{Stecura1978EffectsOC,cao_vassen_stoever_2004,Vasen2022-oq}). However, this composition is limited to a temperature of 1200$^oC$ due to the formation of tetragonal and cubic phases above this temperature, leading to undesired expansion of about 3 to 5 $\%$ (\cite{cao_vassen_stoever_2004, liu_HEC_2022, YSZ_fail_mechanisms_Pakseresht_2022}). It is also susceptible to chemical attack by calcia-magnesia-alumino-silicate (CMAS). To develop a replacement that outperforms $YSZ$ as a coating, the new desirable material would have lower thermal conductivity, comparable fracture toughness ($1-2$ $MPam^{1/2}$), thermal expansion coefficient ($\sim10-11\times10^{-10}$ $K^{-1}$)(\cite{vassen_TBC_overview_2010}) and thermal cyclic life-span, higher resistance to CMAS and better phase stability at its operating temperatures ($\sim1200-1300$ $^oC$)(\cite{bakan_vaßen_2017, Vasen2022-oq, mehboob_TBC_failure_mechanisms_2020}).

Due to the existing limitations of $YSZ$ coatings, there has been an increasing interest in the discovery of high-performance alternatives (\cite{Vasen2022-oq}). Pyrochlore structures are currently the most investigated group (\cite{bakan_vaßen_2017, vassen_zirconates_TBC,fergus_zirconia_pyrochlore_TBC_2014, wu_wei_low_TC_RE_Zirconates}). This is partly due to their lower thermal conductivity, higher thermal stability and lower oxygen diffusivity (\cite{vassen_zirconates_TBC, TC_zhu_meng_zhang_li_xu_reece_gao_2021}) compared to YSZ. However, these compositions have a much lower thermal expansion coefficient and mechanical toughness compared to $YSZ$, making them unsuitable for replacement. The pyrochlore structure can be modified by cation substitution, forming high-entropy or multi-component ceramics (HECs and MCCs). The cubic pyrochlore structure has the general formula $A_2^{3+}B_2^{4+}O_6O'$, where $A^{3+}$ is typically a rare-earth element and $B^{4+}$ is usually a transition metal. By including multiple elements in different combinations in the $A$ and $B$ sub-lattices, both the thermal and mechanical properties of the material can be manipulated (\cite{CWAN_LGZ, ren_LaYb_ReZr_HEC_2015, liu_thermal_prprt_LGZ_2014}). The introduction of elements with different ionic radii, mass and electro-negativities, provides the potential to tailor the thermal and mechanical properties of the material, allowing for an ensemble of properties that would not otherwise be achievable with single elements. Additionally, distortion in the lattice caused by different $Re-O$ bond lengths has been shown to reduce thermal conductivity (\cite{Wright2019SizeDA, tsai_HEA_2014}). The ability to tailor both the mechanical and thermal properties of these compositions makes it a worthwhile search space for potential new TBC materials.

Due to the large number of combinations this allows, experimentally measuring the properties of these potential TBC materials would be too costly and time-consuming. Computational methods using first-principles calculations, such as density functional theory (DFT) take time to set up and run (\cite{Seko_ML_descriptors_materials_discovery_2018}). And third order problems, such as thermal conductivity are computationally expensive and often limited to crystals with a small number of atoms in their unit cell (\cite{Choi_Lee_lattice_TC_NN_2022, Luo_Machine_learning_TC_review_2023}). As an alternative to overcome laborious, resource intensive processes, machine learning (ML) has been increasingly used in materials science, for property and micro-structure predictions, and composition generation (\cite{liu_Materials_Discovery_2017}). Conventional machine learning methods used include random forest, support vector machines and shallow neural networks. They often take as input hand-crafted and/or non-problem specific descriptors to represent information about the input data points representing materials.  These descriptors are hypothesised to be relevant, that is, to have some degree of correlation to the target variable (\cite{Ward2016-lj_General_purpose_ML_material_property_prediction, Jha2018-ig_Deep_learning_with_elemental_composition}). The definition and selection of these descriptors generally requires domain knowledge--- although often generic chemistry descriptors are also used --- and an understanding of the machine learning algorithms (\cite{Ward2016-lj_General_purpose_ML_material_property_prediction, Jha2018-ig_Deep_learning_with_elemental_composition, guyon_feature_selection_intro_2003}). This is an important step, as relevant descriptors can help understand materials structure-property relationships.

Yang \textit{et al.} conducted an investigation of the atomic parameters that mostly affect the prediction of thermal and mechanical properties of pyrochlores {\cite{yang_mechanical_thermal_properties_RE_pyrochlores_2018}).  Parameters include the average atomic mass of $A$ and $B$ cations ($M_A$ and $M_B$), period ($P_A$ and $P_B$) in the periodic table, average cationic radius ($R_A$ and $R_B$), cationic radius ratio ($R_A/R_B$), electronegativity ($EN_A$ and $EN_B$), density ($\rho$) and lattice constant ($\alpha$).  \cite{Miyazaki_ML_TC_prediction_half-heusler_2021} developed a ML model to predict the lattice thermal conductivity of half-heusler compounds using a two stage system, that first predicted lattice parameters using ML and then used the predicted lattice parameter as a descriptor for training the thermal conductivity prediction model. The descriptors for the lattice parameter prediction model included the average atomic radius and average atomic mass of each site in the lattice ($r_1, r_2, r_3, m_1, m_2, m_3$). The thermal conductivity model used the previously predicted lattice parameter and 54 other descriptors. These 54 descriptors were various functions of $r_1, r_2, r_3, m_1, m_2, m_3$ describing crystallographic properties specific to half-heusler compounds. Using the Wrapper method (\cite{Kohavi1997-ni_wrappers_feature_selection}), these 55 descriptors were then reduced to just 4 in order to optimise the training of the ML model. Their database was obtained from DFT calculated values of lattice parameter and thermal conductivity of 143 materials. The predictive models employed were multiple linear regression and boosted decision tree regression, with the latter achieving better results in both stages. For thermal conductivity, the boosted decision tree model performed best with an error of $\pm4\%$ and $R^2$ of 0.84. The final trained model  predicted thermal conductivity near instantaneously with a an accuracy of $\pm 4\%$, while DFT calculations for the same compound took up to 72 h.

Despite the advantages over numerical methods, ML are restricted by their training data and how the data is represented --- that is, the quality of the descriptors defined. ML is expected to learn from data sets that provide a good representation of the search space to achieve good predictive power. Furthermore, trained models have been shown to be inaccurate when predicting properties of compounds with values of the target property that are significantly different to those represented in the training set. A general model trained on various different types of compounds and materials would be ideal, but is difficult to implement with algorithms like Random Forest. Crystal structures are not well represented for modelling when using solely descriptors such as their chemical formula or simple crystal structure information. The position of atoms in the lattice and bonding between atoms in different sites, for instance, have significant effects on thermal transport through the material. Thermal conductivity in crystals, is governed by phonon (or quantised lattice vibration modes) scattering within the lattice~(\cite{pala_thermal_conductivity_phonon_transport_2012}). Phonon scattering is dictated by the strengths and positions of the various bonds present throughout the lattice structure~(\cite{liu_LZ_structure_and_TC_2007}). A representation that captures this information will likely lead to models with more predictive power. In the current literature, most ML models investigating crystal properties use simple descriptors (composition and crystal structure information), being limited to the same type of materials (\cite{Luo_Machine_learning_TC_review_2023, Wright2019SizeDA}).

Graph Neural Networks (GNN) capture structural information as a graph composed of nodes and edges. Edges represent connections between nodes; they also contain edge features information, such as the distance between two nodes. GNNs are commonly used in pharmaceutical research to represent 2D molecules (\cite{carracedo_review_machine_learning_drug_discovery_2021}) for QSAR modelling and materials discovery (\cite{Zhang_GNN_drug_discovery_2022, Jiang_descriptor_vs_graph_representation_2021}). While various methods exist to represent organic molecules as graphs, like SMILES, they are not suitable for representation of 3D, periodic crystal structures. Xie and Grossman, 2018, represented crystals as a graph structure allowing the use of over 3000 data-points to train various property prediction models using their custom algorithm, the Crystal Graph Convolutional Neural Network (CGCNN) (\cite{CGCNN_xie_2018}). They convert crystallographic information files (CIF) to a 2D graph used as the input for CGGNN. \cite{Zhu_CGCNN_Inorganic_Crystal_TC_2021} used CGCNN to successfully predict the thermal conductivity of all known rare earth chalcogenides in the  Inorganic Crystal Structure Database (ICSD), for applications in thermo-electric power generation.

For an improved screening of potential materials for TBC's, different representation methods for high-entropy pyrochlore compositions are explored in this study. We evaluate hand-crafted descriptors and a graph-based representation of the lattice structure. The hand-crafted representations tested use compositional descriptors and  crystallographic descriptors. Compositional descriptors to create a predictive model allow us to study the effect of different atomic species on the thermal conductivity, making it useful for material design. Crystallographic descriptors, however, offer a more standardised representation of materials, potentially reducing model bias due to the distribution of species in the training dataset. Graph-based representation  combines the compositional and crystallographic hand-crafted properties and allows for additional material information to be included. Using interpretative ML, each representation can offer different insights into those properties to be selected in next generation rational designs. This study focuses on the accuracy of the models trained with each approach to determine their potential usefulness both as a predictive model for high-throughput screening, and an interpretative model that can be used to study the science of thermal conductivity. Materials with a pyrochlore structure are the primary focus due to their potential for low thermal conductivity and capability to form high-entropy compositions which provides a sufficiently large search-space to test the capabilities of machine learning. Thermal conductivity is selected as the prediction target as without a sufficiently low thermal conductivity the material would not be able to function as a thermal insulator(\cite{Vasen2022-oq}). For the ML models, the hand-crafted descriptors are used to train Random Forest Regressor (\cite{scikit-learn}) and Gaussian Process Regressor (\cite{scikit-learn}), while the graph based representation was used to train the CGCNN. Training data for the model was gathered from literature.

It should be noted that most of the CIF files used in the training data are not generated by experimental measurements or DFT. DFT calculated CIF files of the multi-component/high-entropy compositions would likely have more accurate information regarding atomic coordinates and bond lengths. This avenue would, however, require large amounts of time and computational resources and was, thus, not explored for this paper.

\section{Methodology}\label{Method}

\subsection{Database}\label{Database}

Thermally sprayed coatings have varying micro-structures that depend on the processing conditions and compositions. To simplify the selection of descriptors, only data from pellets were considered. While DFT generated data was considered, there is a lack of sufficient calculated thermal conductivity data on rare-earth pryochlores, especially high-entropy compositions. To ensure consistency, the data was collected from a single source. The data selected for our investigation was obtained from  \cite{TC_wright_wang_ko_chung_chen_luo_2020}. The authors studied thermal and mechanical properties of 22 primarily single-phase, rare-earth pyrochlores with multiple components in both \textit{A} and \textit{B} positions. The \textit{A} position was composed of rare-earths in various combinations from single element to up to 7 elements. The \textit{B} position was composed of $Zr$, $Sn$, $Hf$ and $Ti$ in various combinations from single element to up to 4 elements. 

Densities of samples were measured using the Archimedes method, following the ASTM C373-18 Standard. Relative density of the samples are around 95\% on average .Thermal conductivity was calculated from thermal diffusivity which was measured at room temperature. Thermal diffusivity was measured using LFA 467 \textit{HT Hyperflash} (NETZCSH, Germany), laser flash analysis. Specific heat was calculated using Neumann-Kopp (\cite{suresh_TC_LZ_GZ, leitner_Neumann-Kopp_application_2010}) rule and used to calculate the thermal conductivity of the pellets.

\subsection{Hand-Crafted Descriptors}

\textit{Yang et al.} used least absolute shrinkage and selection operator (LASSO) to investigate atomic/crystallographic parameters affecting the prediction of thermal and mechanical properties of pyrochlores (\cite{yang_mechanical_thermal_properties_RE_pyrochlores_2018}).  Parameters include atomic mass of $A$ and $B$ cations ($M_A$ and $M_B$), period ($P_A$ and $P_B$) in the periodic table, ionic radius ($R_A$ and $R_B$), cationic radius ratio ($R_A/R_B$), electronegativity ($EN_A$ and $EN_B$), density ($\rho$) and lattice constant ($\alpha$). Compositional parameters, represented by atomic fractions of each element present in the composition, were considered  alongside crystallographic properties, such as the effective mass and radii of each cation, as well as the lattice parameter. Table \ref{Table_model_features} lists all descriptors considered for our model.

%m{5cm}ccc

\begin{table}[H]
\begin{center}
\begin{tabular}{cm{4.35cm}} 
 \hline
 Descriptors & Description\\
 \cmidrule(lr){1-1}\cmidrule(lr){2-2}
 $LaA - YbA$  & Atomic fraction of elements in the $A$ cation site, \\
              & from lanthanum to ytterbium, normalised to 1.  \\
 \cmidrule(lr){2-2}
 $ZrB, HfB, SnB, CeB,$ & Atomic fraction of elements in the $B$ cation\\
 $TiB$ & site, normalised to 1.\\
 \cmidrule(lr){2-2}
 $RA$ & Effective ionic radii of $A$ cations\\
 \cmidrule(lr){2-2}
 $RB$ & Effective ionic radii of $B$ cations\\
 \cmidrule(lr){2-2}
 $MA$ & Effective atomic mass of $A$ cations\\
 \cmidrule(lr){2-2}
 $MB$ & Effective atomic mass of $B$ cations\\
 \cmidrule(lr){2-2}
 $RA/RB$ & Cationic ratio\\
 \cmidrule(lr){2-2}
 $P$ & Presence or absence of pyrochlore phase represented by 1 or 0.\\
     & This was based on the $RA/RB$ criteria for pyrochlore. \\
 \cmidrule(lr){2-2}
 $a$ & Theoretical lattice constant calculated using (\cite{mouta_pyrochlore_lattice_constant_2013}).\\
 \cmidrule(lr){2-2}
 $Entropy$ & Lattice configurational Entropy\\
 \hline
\end{tabular}
\caption{ \label{Table_model_features}Brief description of the 26 descriptors used to describe the compositions in the model}
\end{center} 
\end{table}

Theoretical lattice constant ($a$) was calculated using Equation \ref{eqtn_mouta_lattice_constant} from Mouta (\cite{mouta_pyrochlore_lattice_constant_2013}) using the average ionic radii of the cations.

\begin{equation}
    a_2 = \frac{8}{3^{1/2}}\left[1.43373(R_A + R_O) - 0.42931\frac{(R_A + R_B)^2}{R_B + R_O} \right]
    \label{eqtn_mouta_lattice_constant}
\end{equation}

Where $R_A$, $R_B$ and $R_O$ are the ionic radii of the $A$ cation, $B$ cation and Oxygen, respectively. This was used as a quick method to obtain the lattice parameters of multi-component composition data obtained from literature.

Stability of the pyrochlore structure is dependant on the ratio of ionic radii of $A$ and $B$ cations ($R_A/R_B$). For lanthanide zirconates ($ln_2Zr_2O_7$), when $R_{A}/R_{B}\leq1.46$ (\cite{fuentes_pyrochlore_ra_over_rb_criteria_2018}) a flourite phase is formed instead. A simple phase descriptor ($P$) is used to differentiate between the two phases. The remaining descriptors are applicable to both pyrochlores and fluorites. This is possible because the database used consists mainly of single-phase compositions.

The configurational entropy refers to the portion of the entropy of a system related to the precise positions of individual particles that make up the system. This parameter accounts for the number and proportion of each distinct atomic species present in the $A$ and $B$ sub-lattices of the pyrochlore lattice. To account for multiple sub-lattices in complex crystal structures,  (\cite{Dippo_config_entropy_2021}) proposed a new entropy metric shown in Equation \ref{eqtn_dippo_entropy_metric}.

\begin{equation}
    EM = \frac{S_{SL/\text{mol atoms}}^{config}}{R}*L
    \label{eqtn_dippo_entropy_metric}
\end{equation}

Where $R$ is the ideal gas constant, $L$ is the total number of sub-lattices and $S_{SL/mol_atoms}$ is the configurational entropy calculated using the sub-lattice model shown in Equation \ref{eqtn_sublattice_entropy}.

\begin{equation}
    S_{config}^{SL} = \frac{-R\sum_S\sum_ia^SX_i^S\ln{(X_i^S)}}{\sum_Sa^S}
    \label{eqtn_sublattice_entropy}
\end{equation}

Where $a^S$ is the number of sites on the $S$ sub-lattice and $X_i$ is the fraction of each individual species $i$ in the sub-lattice. For the $A_2B_2O_6O'$, $\sum a_S$ is 11 and $L$ is 5 when considering the oxygen vacancy as an additional species.

Given that all the crystallographic parameters are directly calculated from the atomic fractions, it would be redundant to consider them together. A correlation matrix (figure \ref{Img_Corr_matrix}) further supports this argument. Due to that, the atomic fractions and the atomic/crystallographic parameters are used as two separate methods. The crystallographic parameters are a more common representation of the problem than the atomic fractions. However, the effect of composition on thermal conductivity is also of interest, as it would be more intuitive from a materials design perspective.

\subsubsection{Machine Learning Algorithms for hand-crafted descriptors}\label{Sec_Model}

Random Forest (RF) (\cite{scikit-learn}) and Gaussian Process (GP) regressors (\cite{scikit-learn}) were selected to create predictive models for our data. They were implemented using the the \textit{scikit-learn} package (\cite{scikit-learn}) in \textit{Python} 3.9.12. Hyper-parameters were optimised using\textit{GridSearchCV}.

The RF was set up with 1000 decision trees (\textit{n\_estimators}) and a value of 2 was used for \textit{min\_sample\_leaf} based on the results using \textit{GridSearchCV}. Random state of the model was set to 42. All other values were kept at default for the package.

The GP used a combination of \textit{ConstantKernel} and \textit{Radial Basis Function} (RBF) for the kernel. For \textit{ConstantKernel}, the constant was set to 1.0 and the bounds were set to ``fixed". For \textit{RBF}, length scale was set to 1.0 and the bounds were also set to ``fixed".

\subsection{Graph-based representation} \label{Sec_CIF}

The CGCNN takes CIF files representing crystal structures as inputs~(\cite{CGCNN_xie_2018}). The CIF is converted to a graphical representation, with nodes carrying atomic species data and edges denoting inter-atomic bonding with their corresponding distance. A simplified representation is shown in Figure \ref{Img_graph_structure}. 

\begin{figure}
    \centering
    \includegraphics[width=0.5\textwidth]{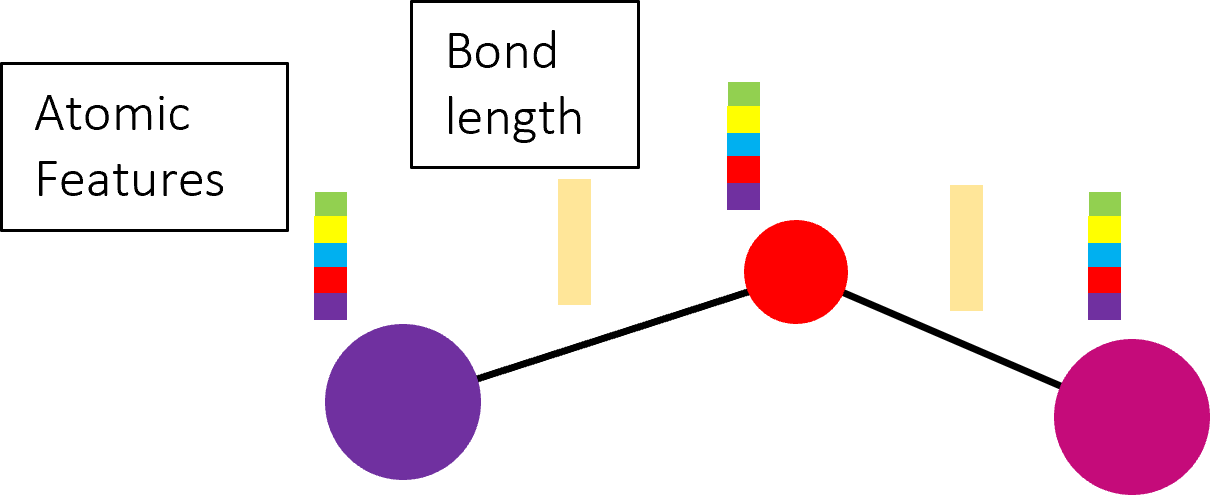}  
    \caption{Structure of the graph used to represent each composition. The nodes represent each atom in the unit cell, containing 9 atomic descriptors and the edges represent bonds between two atoms and contain information about the bond length.}
    \label{Img_graph_structure}
\end{figure}

The nodes contain detailed atomic information about the atomic species represented by the node. Bonded atoms are defined as being located within a radius of 3{\AA} around each atom. Bonds are formed starting with the shortest bond length up to a maximum of 12 possible bonds. The CIF file only contains information about the unit cell; bonded atoms outside the cell are not represented. Due to the symmetry of crystal lattices, this limitation can be overcome by forming a virtual bond with another atom in the equivalent position across the line of symmetry. This allows the periodicity of the crystal to be represented in the graph. Figure \ref{Img_CGCNN_virtual_bond} shows the virtual bond between a lanthanum atom (atom 0) and an oxygen atom (atom 41), across the cell. The inter-atomic distance between the atoms is found by subtracting the actual distance from the lattice parameter.

\begin{figure}
    \centering
    \includegraphics[width=0.5\textwidth]{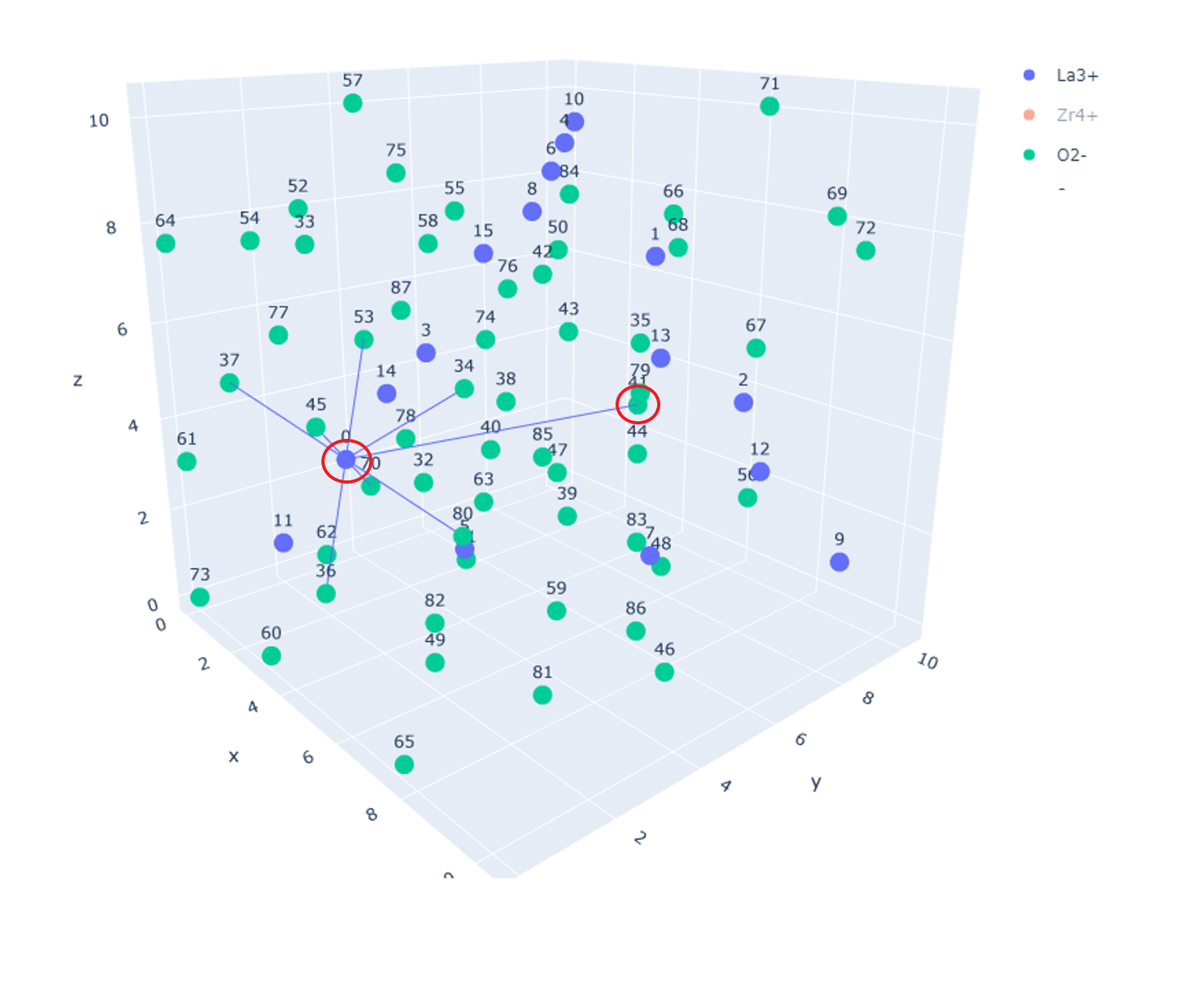}  
    \caption{Virtual bond between a lanthanum atom (atom 0) and an oxygen atom (atom 41), marked by red circles }
    \label{Img_CGCNN_virtual_bond}
\end{figure}

The CGCNN is commonly used with symmetrised CIF files that represent the \textit{occupancy} or probability of an atom species to exist in a particular site within the unit cell, rather than the actual coordinate of each individual atom. For single-element pyrochlores, the occupancy of each species in its specific sub-lattice is 1.0. In high-entropy or multi-component pyrochlores, the species have partial occupancies, with values smaller than 1.0. The CGCNN is unable to handle partial occupancies from the conventionally used symmetrised CIF files. The materials project  (\cite{Jain_Materials_project_al._2013}) possesses an alternative, \textit{computed} CIF file, which contain no symmetry information, but contains the actual Cartesian coordinates of each atom present in the lattice.

\subsubsection{Generating \textit{Crystallographic Information Files}}

While CIF files for the single-element rare earth compositions are easily found in databases such as the Inorganic Crystal Structure Database (\cite{ICSD}) and the Materials Project (\cite{Jain_Materials_project_al._2013}), there are very few available for multi-component pyrochlores. Either experimental measurements or DFT calculations are required to generate the lattice constants of different crystals. This method is very time-consuming to generate the CIFs for the all data being used in this project. CIF files for multi-component compositions were generated artificially by using an existing pyrochlore CIF file as a template, replacing the atomic species present at each coordinate and changing the lattice parameters. CIF files using the conventional standard, i.e, with the distribution of species represented by sub-lattice occupancies rather than cartesian coordinates, cannot be used to generate the lattice graphs needed for the GNN, specifically in the case of HECs and MCCs. This is because HECs and MCCs contain partial occupancies, i.e, occupancy values that are less than one. As the graph consists of nodes and edges representing the actual position and connections between atoms, it cannot deal with partial occupancies. To account for this, the non-symmetrised format for CIFs are used to generate the lattice graphs. The non-symmetrised formats always use the $P1$ space group and contain real positions for each atom in the lattice, rather than the partial occupancies. For compositions with a fractional number of atoms, i.e, 5.333 atoms of $Zr$, $Hf$ and $Ti$, the values are rounded by randomly sampling the species from the list for the A/B sub-lattices and rounding it up or down as needed to keep the total number of atoms at 16 (the maximum number of atoms in the A and B sub-lattices) and then iterating until all species in the sub-lattice have discrete numbers. Furthermore, the positions of each atom within the appropriate sub-lattice is randomised to avoid clustering of the same species. Detailed steps of the method are shown in Table \ref{Table_CIF_generation}. Figure \ref{Img_HEC_CIF} shows the visualisation of the unit cell of $(Sm_{1/3}Gd_{1/3}Eu_{1/3})_2(Zr_{1/2}Hf_{1/2})_2O_7$, generated by our method. In this case, the europium atoms in the $16d$ site (A sub-lattice), are replaced with samarium, gadolinium and europium according to their proportions.

\begin{table}[H]
\begin{center}
\begin{tabular}{cm{7cm}} 
 \hline
  & CIF generation algorithm\\
 \hline
 1: & Read template CIF \\
 \cmidrule(lr){2-2}
 2: & Read database csv file\\
 \cmidrule(lr){2-2}
 3: & Iterate over each entry in database\\
 \cmidrule(lr){2-2}
 4: & Create new file for each entry and copy template data into the new file\\
 \cmidrule(lr){2-2}
 5: & Replace lattice parameter, cell volume, composition on appropriate lines using 
 info from database\\
 \cmidrule(lr){2-2}
 6: & Calculate number of atoms in A and B sub-lattices using atomic fractions of species in database entry\\
 \cmidrule(lr){2-2}
 7: & Round the number of atoms (for HECs and MCCs) ensuring the total number of atoms is 16 for each sub-lattice\\
 \cmidrule(lr){2-2}
 8: & Create two separate lists of all atoms in A and B sub-lattices and shuffle the order of atoms in each list to randomise their positions within each sub-lattice\\
 \cmidrule(lr){2-2}
 9: & Go through the A/B sub-lattice coordinates in the new file, replacing each species present in that coordinate with the species from the aforementioned list \\

 \hline
\end{tabular}
\caption{ \label{Table_CIF_generation}Method for generating CIF for high-entropy and multi-component ceramics}
\end{center} 
\end{table}

\begin{figure}
    \centering
    \includegraphics[width=0.5\textwidth]{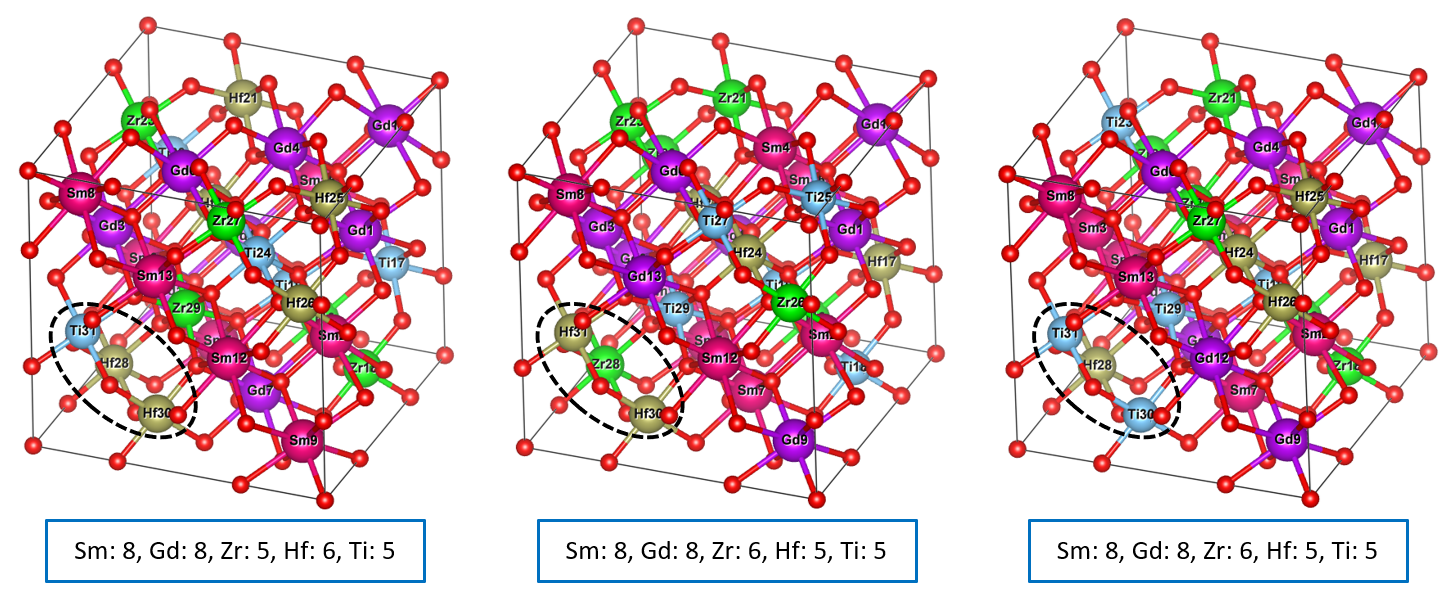}  
    \caption{Three iterations of $(Sm_{1/3}Gd_{1/3}Eu_{1/3})_2(Zr_{1/2}Hf_{1/2})_2O_7$ with slightly differing proportions of each species (as shown below each image) and positions of atoms within the same sub-lattice. Here the changes are only in the $B$ sub-lattice as the number of $Sm$ and $Gd$ atoms are exactly 8.}
    \label{Img_HEC_CIF}
\end{figure}

The template file used for generating the CIFs is $Eu_{2}Zr_{2}O_{7}$, obtained from the Materials Project~(\cite{MP_Eu2Zr2O7_template_2020}). Lattice parameters and cell volumes were obtained using Equation \ref{eqtn_mouta_lattice_constant}. 

\subsection{Data Augmentation}

Data augmentation uses existing data to artificially create new instances for training ML. This is particularly useful for small data-sets, where the augmented data can improve both  size and data distribution. Two augmentation techniques were applied to the data, one for the handcrafted descriptors and another for the graph-based descriptors. The objective was to further test the efficacy of the input data representation for more points, other than those provided in the original data investigated.

\subsubsection{Handcrafted Descriptors - SMOGN}

For the handcrafted descriptors, a data balancing algorithm, namely SMOGN (\cite{smogn}) was applied to augment the dataset. SMOGN  combines over-sampling of minority cases with Gaussian noise to generate synthetic data points.   It starts by identifying regions where the target variable is under represented (minority cases). Subsequently, it generates synthetic data points in these regions. Synthetic data are created by interpolating nearby instances within the minority region. Gaussian noise is then added to the synthetic instances to ensure that the new points are not simply linear combinations of existing data. This helps to improve generalisation and avoid potential biases in the model. Areas with more scarcity of available instances are identified via a relevance value between 1 and 0 (1 being higher relevance). The default function for identifying relevant regions for over-sampling assumes a bell curve distribution of values in the data-set, attributing higher relevancy to values on either end of the distribution. To prioritise increasing the number of data points available, we define the relevancy points as follows (Table \ref{Table_relevancy_mtrx}):

\begin{table}[H]
\begin{center}
    \begin{tabular}{cc} 
         \hline
         Thermal conductivity\\($W/mK$) & Relevancy\\
         \cmidrule(lr){1-1}\cmidrule(lr){2-2}
         1.36 & 1\\	         
         1.5	& 1\\	         
         1.7	& 0.8\\
         1.97 & 0.6\\
         2.1 & 0.8\\  
         2.145 & 1\\
         2.29 & 1\\
         \hline
    \end{tabular}
\caption{Custom relevancy matrix for SMOGN}
\label{Table_relevancy_mtrx}
\end{center} 
\end{table}

It should be noted, that SMOGN is mainly designed to balance an existing data-set and does not simply increase the number of data points by an arbitrary amount. The variable chosen for optimisation was thermal conductivity. A total of 13 new, synthetic data points were generated with SMOGN. Of those 13, 4 were found to be duplicates. Overall, 9 new data points were generated through SMOGN, increasing the database size from 21 to 30. The principal component analysis plot (PCA), of the SMOGN augmented data-set is showin in Figure \ref{Img_SMOGN_PCA}.

\begin{figure}
    \centering
    \includegraphics[width=0.5\textwidth]{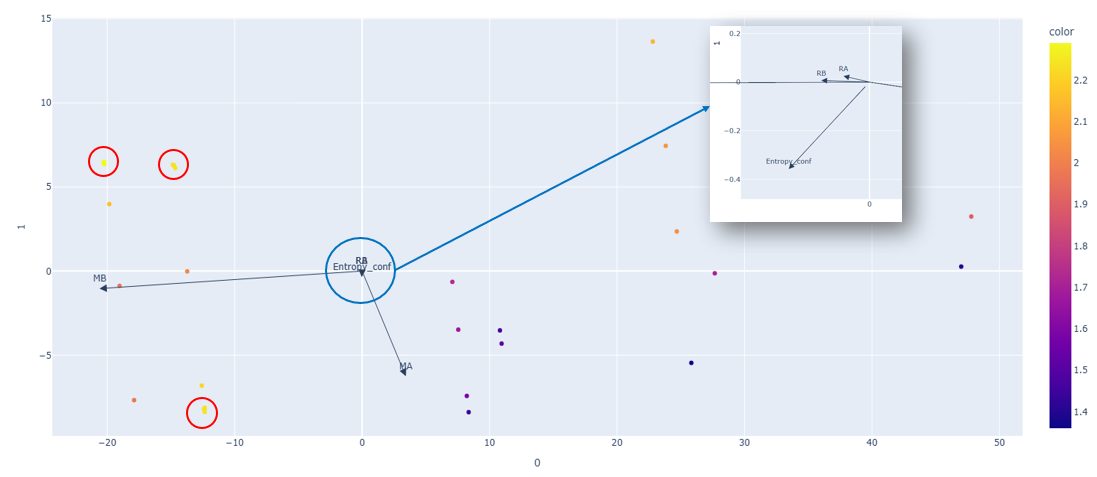}  
    \caption{PCA plot of SMOGN augmented data-set}
    \label{Img_SMOGN_PCA}
\end{figure}

The figure shows a correlation between high $M_B$ and low $M_A$ with high thermal conductivity. The data points added by SMOGN are clustered within the areas denoted by red circles.

The relevancy matrix was designed to generate the most number of data points, which is likely why the distribution of TC for the augmented data-set is still skewed and even slightly worse than the original. The majority of the generated data points had a TC value of approximately 2.2 $W/mK$. A cursory validation of the generated data points was done by using a simple script to match potential compositions of $A_{2}B_{2}O_{7}$ structure to the descriptors. The thermodynamic stability of these compositions were not validated theoretically or experimentally as this is outside the scope of this paper. 

\subsubsection{Graph-based Data - Lattice perturbation}

Multi-component crystal lattices can contain random variations in the positions and amount of different species, within their respective sub-lattice (\cite{Lattice_randomness_Aidhy_2024}). These variations were used to generate copies of each composition in the database, tripling the size of the initial training set. Figure \ref{Img_HEC_CIF} shows three versions of the same composition ($(Sm_{1/3}Gd_{1/3}Eu_{1/3})_2(Zr_{1/2}Hf_{1/2})_2O_7$) with slight differences in the position and proportion of each species. The difference in proportion arises due to fractional occupancy.

\subsection{CGCNN Model}

The CGCNN architecture consists of 3 convolutional layers, 2 hidden layers and 1 pooling layer. Each convolution layer, learns the information from neighbouring atoms to generate a new feature vector for each atom in the graph. Following that, the pooling layer then uses the information from the convolutional layers to generate a vector represeting the entire crystal.

\begin{figure}
    \centering
    \includegraphics[width=0.5\textwidth]{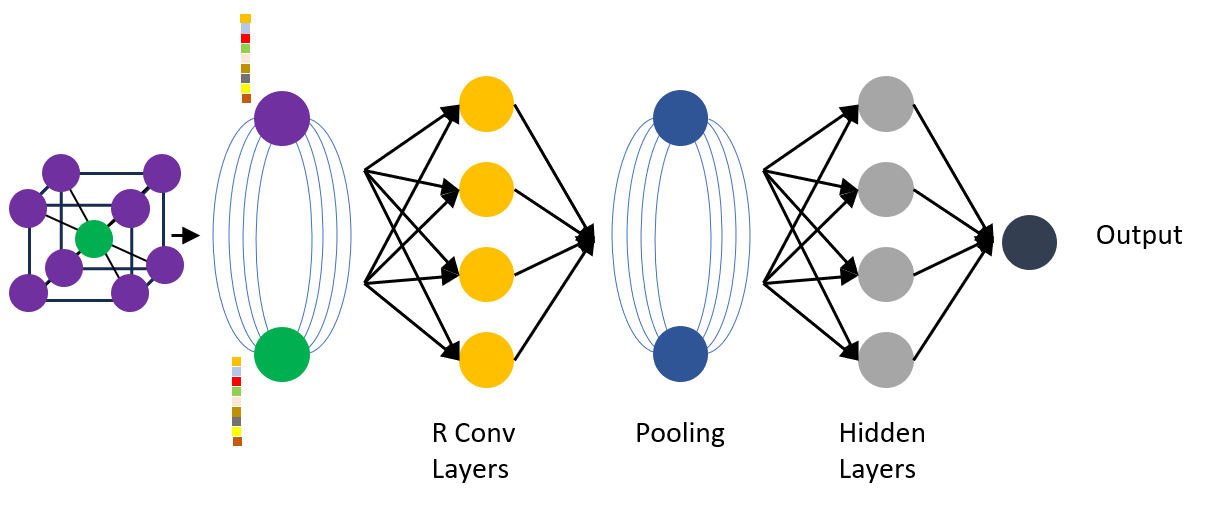}  
    \caption{CGCNN model architecture. CIF file is converted to a graph consisting of nodes and edges representing atoms and their bonds, respectively. The nodes are embedded with vectors containing information representing the atomic species which are then aggregated in the pooling layer to represent the entire crystal}
    \label{CGCNN_architecture}
\end{figure}

The mean squared error is used as the loss function during training. Table \ref{Table_hyperparameters_CGCNN} shows the hyperparameters used for the CGCNN model.

\begin{table}[H]
\begin{center}
    \begin{tabular}{cc} 
         \hline
         Hyper-Parameter  & Value \\
         \cmidrule(lr){1-1}\cmidrule(lr){2-2}
         learning rate & 0.01\\
         batch-size & 64\\
         Optimiser & SGD\\
         \hline
    \end{tabular}
\caption{Effect of element distribution on prediction}
\label{Table_hyperparameters_CGCNN}
\end{center} 
\end{table}

It should be noted that the bond length is not included as a descriptor for the GNN. Calculating the bond lengths of different $Re-O$ pairs in HECs and MCCs is a complicated process that would require DFT modelling for each composition. This is outside the scope of the current work and the decision was made to remove bond lengths as a descriptor entirely.

\subsection{Model Evaluation}

. Mean Squared Error (Equation \ref{eqtn_MSE}) is used for the CGCNN during the training process. Ultimately, mean absolute error in Equation \ref{eqtn_MAE} is used to evaluate all the models as it is more intuitive than the MSE.

\begin{equation}
    MAE = \frac{1}{N}\sum_{i=1}^n\abs{y_i - y^*_i}
    \label{eqtn_MAE}
\end{equation}
\begin{equation}
    MSE = \frac{\sum_{i=1}^N(y_i - y^*_i)^2}{N}
    \label{eqtn_MSE}
\end{equation}
\begin{equation}
    R^2 = 1 - \frac{\sum_{i=1}^N(y_i - y^*_i)^2}{\sum_{i=1}^N(y_i - \Bar{y})^2}
    \label{eqtn_R2}
\end{equation}

Where, $y_i$ is the actual measured value, $y^*_i$ is the predicted value and $\Bar{y}$ is the mean of the actual values. Leave-one-out cross validation (LOOCV) is used to evaluate all models due to the limited size of the data-set. The MAE from each fold is then averaged and used to compare the different models as it is a more intuitive metric.The $R^2$ was calculated by combining the results of all folds, for each model.

\section{Results}\label{Results}

A summary of the model error metrics derived from LOOCV is listed in Table \ref{Table_model_comparison_errors_LOOCV}, while Table \ref{Table_model_comparison_errors_LOOCV_SMOGN} lists the performance of the models trained using the augmented datasets (both SMOGN and Lattice perturbation). 

\begin{table}[!H]
\begin{center}
  \begin{tabular}{lccc}
    \hline
    \  &
      \multicolumn{2}{c}{MAE} &
      \multicolumn{1}{c}{$R^2$} \\
    \hline
    Model & Mean & Std. Dev & Mean\\
    \cmidrule(lr){1-1}\cmidrule(lr){2-2}\cmidrule(lr){3-3}\cmidrule(lr){4-4}
    (1) GP (compositional) & 0.209 & 0.245 & -1.255\\
    (2) RF (compositional) & 0.117 & 0.084 & 0.776\\
    (3) RF (crystallographic) & 0.112 & 0.089 & 0.781\\
    (4) CGCNN & \textbf{0.029} & \textbf{0.043} & \textbf{0.97}\\
    \hline
  \end{tabular}
  \caption{Leave-One-Out-Cross validation results for models trained on the original dataset}
  \label{Table_model_comparison_errors_LOOCV}
\end{center}
\end{table}

Figures \ref{Img_GP_comp_LOOCV_pred-actual_plot}-\ref{Img_CGCNN_Aug_LOOCV_pred-actual_plot} show the predicted vs. actual thermal conductivity values for each composition, using the different models. The Gaussian Process model trained with compositional descriptors shows the worst performance in all metrics. The negative $R^2$ value indicates that the model was unable to learn any meaningful relationships within the data. The model performance seems to be strongly affected by a single data point ($Sm_2Zr_2O_7$). As seen in Figure \ref{Img_GP_comp_LOOCV_pred-actual_plot}, this compositon has a significantly higher error than the average ( 55 $\%$ vs the average of 11 $\%$). The predicted thermal conductivity is significantly lower than the minimum in the training data-set. This point could be considered an outlier in the sense that it is one of only two single-element compositions in the training data, the other being lanthanum zirconate ($La_2Zr_2O_7$). RF trained with compositional descriptors shows improvement over the GP model. Interestingly, the differences between the RF models using compositional and crystallographic descriptors seem to be minimal. Both models also show a similar trend in the error values for each composition as can be seen in figures \ref{Img_GP_comp_LOOCV_pred-actual_plot} and \ref{Img_RF_comp_LOOCV_pred-actual_plot}. This implies that both sets of descriptors are similarly informative in regards to thermal conductivity. The Random Forest model trained on the SMOGN augmented data-set shows a significant improvement over the regular data-set. However, as can be seen from figure \ref{Img_RF_SMOGN_crys_LOOCV_pred-actual_plot}, the error trend for each composition appears to be similar. Particularly, in both models, the composition $(Sm_{1/2}Gd_{1/2})_2\allowbreak(Zr_{1/3}Hf_{1/3}Ti_{1/2})_2O_7$ exhibit a higher error compared to other data points in the region. The augmented model appears to work better for thermal conductivity between 2.0 and 2.4 $W/mK$. The $R^2$ is also significantly improved, suggesting better generalisability compared to model 3.

\begin{table}[h]
\begin{center}
  \begin{tabular}{lccc}
  \hline
    Model & RF (compositional) & RF (crystallographic) & CGCNN\\
    \cmidrule(lr){1-1}\cmidrule(lr){2-2}\cmidrule(lr){3-3}\cmidrule(lr){4-4}
    GP (compositional) & 0.288 & 0.128 & $8.39\times{}10^{-5}$\\
    RF (compositional) & - & 0.973 & $1.2\times{}10^{-3}$\\
    RF (crystallographic) & - & - & $1.6\times{}10^{-3}$\\
    \hline
  \end{tabular}
  \caption{Statistical significance test to study differences between models using Wilcox Signed Rank Test}
  \label{Table_model_comparison_Wilcox}
\end{center}
\end{table}

\begin{table}[htpb]
\begin{center}
  \begin{tabular}{lccc}
    \hline
    \  &
      \multicolumn{2}{c}{MAE} &
      \multicolumn{1}{c}{$R^2$} \\
    \hline
    Model & Mean & Std. Dev & Mean\\
    \cmidrule(lr){1-1}\cmidrule(lr){2-2}\cmidrule(lr){3-3}\cmidrule(lr){4-4}
    (5) RF(crystallographic,SMOGN) & 0.061 & 0.072 & 0.908\\
    (6) CGCNN (augmented) & \textbf{0.024} & \textbf{0.042} & \textbf{0.975}\\
    (7) CGCNN (SMOGN) & 0.047 & 0.081 & 0.872\\
    \hline
  \end{tabular}
  \caption{Leave-One-Out-Cross validation results for models trained on augmented data}
  \label{Table_model_comparison_errors_LOOCV_SMOGN}
\end{center}
\end{table}

\begin{figure}
    \centering
    \includegraphics[width=0.5\textwidth]{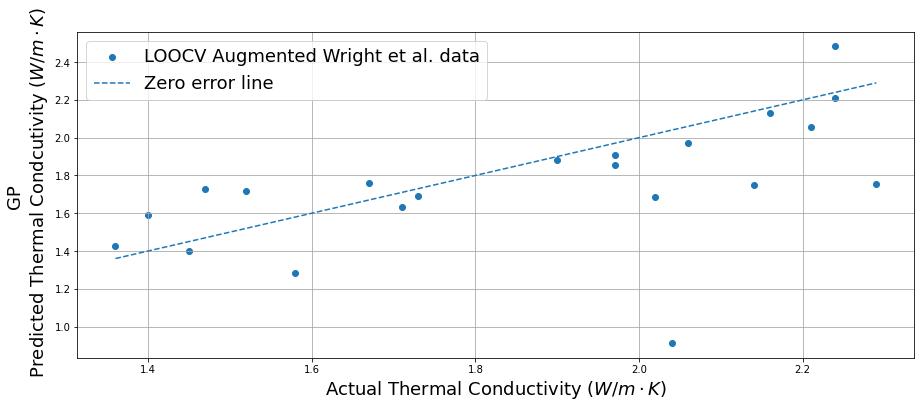}  
    \caption{Predicted vs actual thermal conductivity plot from leave-one-out-cross-validation of GP model using compositional descriptors}
    \label{Img_GP_comp_LOOCV_pred-actual_plot}
\end{figure}

\begin{figure}
    \centering
    \includegraphics[width=0.5\textwidth]{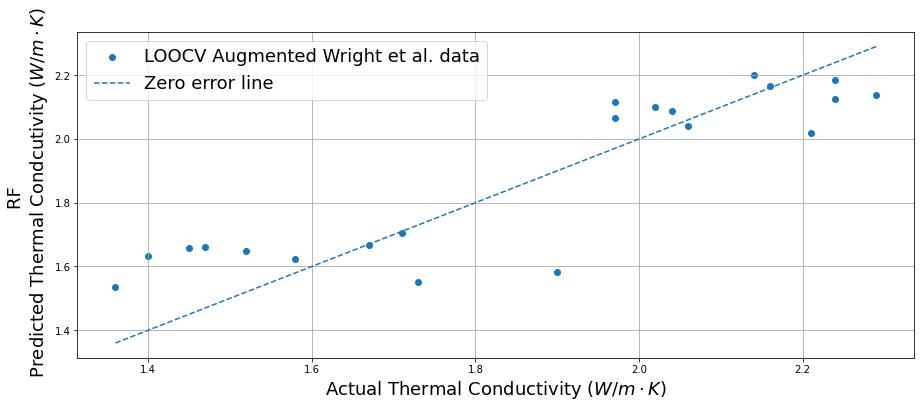}  
    \caption{Predicted vs actual thermal conductivity plot from leave-one-out-cross-validation of RF model using compositional descriptors}
    \label{Img_RF_comp_LOOCV_pred-actual_plot}
\end{figure}

\begin{figure}
    \centering
    \includegraphics[width=0.5\textwidth]{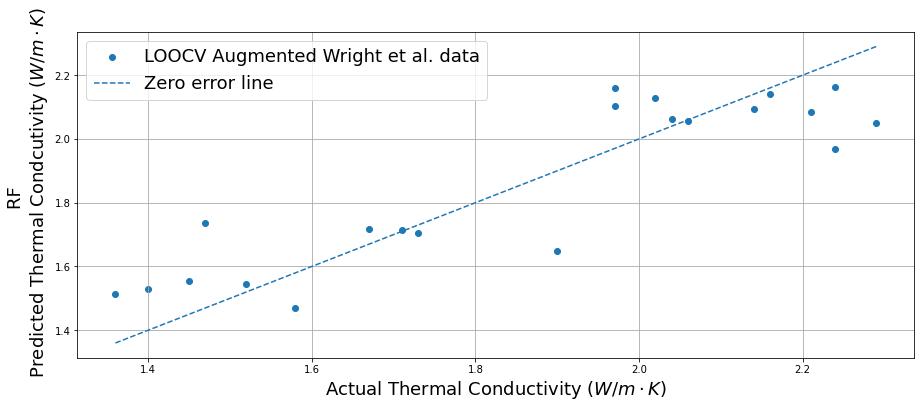}  
    \caption{Predicted vs actual thermal conductivity plot from leave-one-out-cross-validation of RF model using crystallographic descriptors}
    \label{Img_RF_crys_LOOCV_pred-actual_plot}
\end{figure}

\begin{figure}
    \centering
    \includegraphics[width=0.5\textwidth]{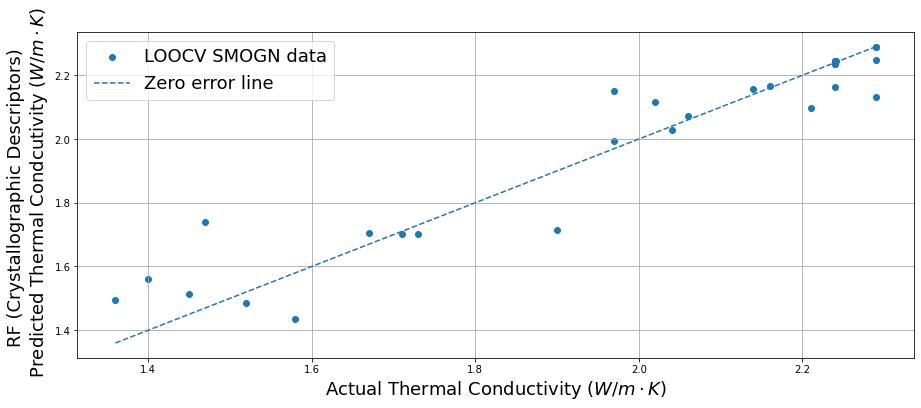}  
    \caption{Predicted vs actual thermal conductivity plot from leave-one-out-cross-validation of RF model trained on SMOGN augmented data-set, using crystallographic descriptors}
    \label{Img_RF_SMOGN_crys_LOOCV_pred-actual_plot}
\end{figure}

\begin{figure}
    \centering
    \includegraphics[width=0.5\textwidth]{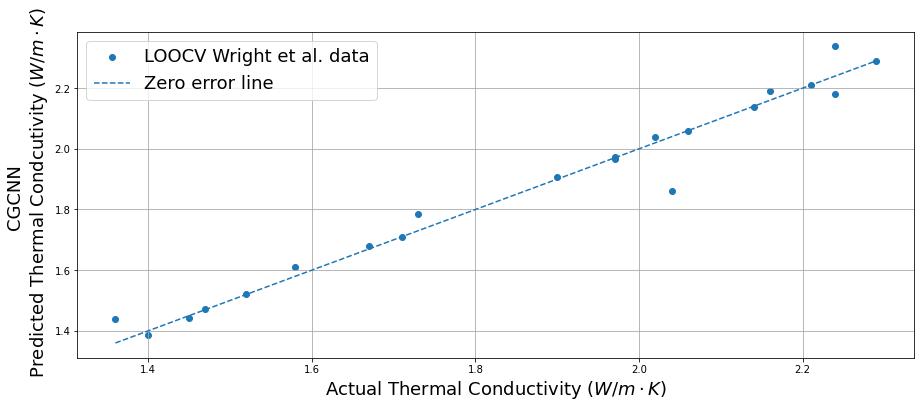}  
    \caption{Predicted vs actual thermal conductivity plot from leave-one-out-cross-validation of CGCNN model}
    \label{Img_CGCNN_LOOCV_pred-actual_plot}
\end{figure}

\begin{figure}
    \centering
    \includegraphics[width=0.5\textwidth]{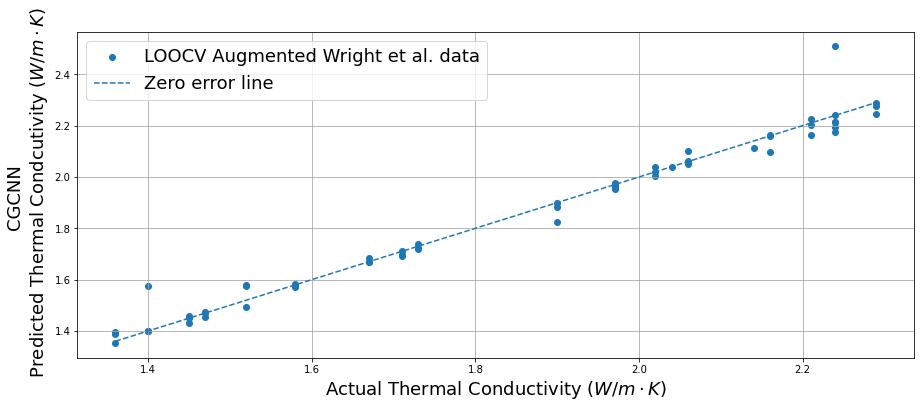}  
    \caption{Predicted vs actual thermal conductivity plot from leave-one-out-cross-validation of CGCNN model trained on the augmented data-set}
    \label{Img_CGCNN_Aug_LOOCV_pred-actual_plot}
\end{figure}

\begin{figure}
    \centering
    \includegraphics[width=0.5\textwidth]{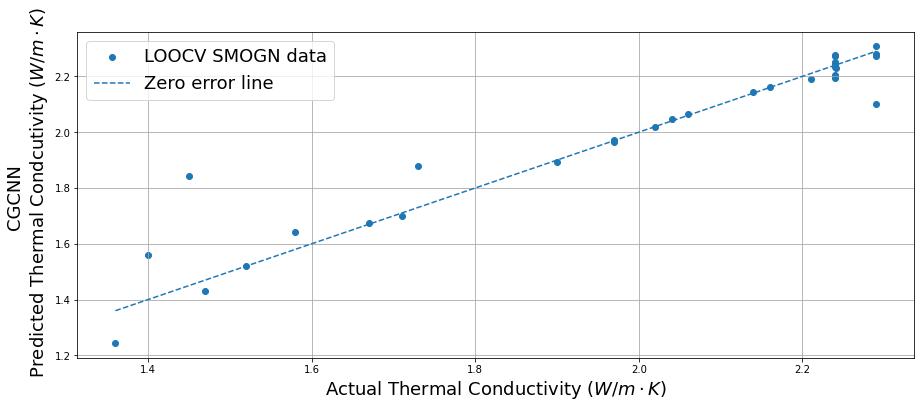}  
    \caption{Predicted vs actual thermal conductivity plot from leave-one-out-cross-validation of CGCNN model trained on the SMOGN data-set}
    \label{Img_CGCNN_SMOGN_LOOCV_pred-actual_plot}
\end{figure}

All three CGCNN models show significantly greater performance compared to the RF and GP models, with the MAE well below 0.1 and $R^2$ exceeding 0.9. The augmented lattice-perturbation data offers a slight improvement in performance. The variation in prediction error for different lattice arrangements of the same composition is minimal, which suggests that the model can account for these slight variations in the lattice. The SMOGN data, when applied to the CGCNN algorithm, decreases the performance compared to model 7 (Figure \ref{Img_CGCNN_SMOGN_LOOCV_pred-actual_plot}), though it is comparable to model 5. The primary contributor to this decreased performance appears to be the composition $(Sm_{1/4}Gd_{1/4}Eu_{1/4}Yb_{1/4})_2\allowbreak(Zr_{1/4}Hf_{1/4}Sn_{1/4}Ti_{1/4})_2O_7$, which has an error of $27.1 \%$ compared to the average error of $2.8 \%$. This data point is from the original data-set. It should be noted, that the compositions generated for the SMOGN data were not a perfect match to the crystallographic parameters, although they were very close. This introduces an additional source of error in the training data. Furthermore, the TC of the generated data were roughly between 2.2-2.4 ($W/mK$). For both the RF and CGCNN algorithms, the additional data appears to have improved the accuracy for this range of thermal conductivity. However, for the CGCNN algorithm, it has also significantly decreased performance in the lower range of thermal conductivity. Also, it should be noted that the TC of the generated data is not experimentally verified. Augmenting the data through lattice perturbation appears to be more effective, leading to a more consistent performance across the full range of TC. However, the latter method is still limited in that it is not able to provide data for sparse regions unlike SMOGN.

Table \ref{Table_model_comparison_Wilcox} shows the results of the Wilcox signed rank test, applied to compare the MAE difference between each model trained on the original data-set. Values below 0.05 indicate a significant difference, while higher values indicate that the difference is of low significance. As stated before, the difference between the RF models using compositional and crystallographic descriptors is minimal. It is likely that a similar level of meaningful information can be gained from both sets of descriptors. The difference between CGCNN and the remaining models is, however, significant. The performance of the CGCNN models could be attributed to the fact that the graph-based representation of the data contains both compositional and crystallographic descriptors, together with the addition of spatial descriptors. Unlike the RF and GP models, the CGCNN models were also able to accurately predict the TC of the singular composition in the region of TC between 1.8 and 1.95 $W/mK$ ($(Sm_{1/3}Gd_{1/3}Eu_{1/3})_2\allowbreak(Zr_{5/6}Hf_{5/6}Sn_{5/6}Ti_{3/4})_2O_7$). This holds promise in the ability of the CGCNN model to extrapolate for regions that are sparse in data.

\section{Conclusions}\label{Conclusions}

This study compared three data representation methods for predicting the thermal conductivity (TC) of multi-component rare-earth pyrochlores; 1) chemical compositional descriptors, 2) simple crystallographic descriptors and 3) graph-based representations encoding the information from crystallographic information files (CIF). The first two were used to train Random Forest (RF) and Gaussian Process (GP) models, while the Crystal Graph Convolutional Neural Network (CGCNN) algorithm was applied to the graph-based representation. Leave-one-out cross-validation (LOOCV) showed CGCNN achieving the best performance with low MAE and high $R^2$, while GP had the worst performance with a negative $R^2$. The RF models trained on compositional and crystallographic descriptors performed similarly suggesting both methods contain comparable information.

To address potential over-fitting from the small dataset, two augmentation techniques were applied. The first, using the SMOGN algorithm, introduced random perturbations in the crystallographic descriptors and TC values. The RF model trained on method 2 showed an increase in performance values using the SMOGN generated data, but the CGCNN showed a reduction in performance.

The second augmentation method introduced small variations in atomic positions and site occupancies in the CIF files, without altering lattice parameters or the composition. GCNN models trained on this dataset showed slight performance improvements with minimal variation in TC predictions across copies. Both augmented and non-augmented CGCN models outperformed SMOGN augmented RF models, achieving $R^2$ values above 0.97. Although this suggests possible over-fitting, both CGCNN models accurately predicted the TC of a composition from the sparsest region of the dataset with less than 1$\%$ error, indicating strong generalizability. In contrast RF and SMOGN-augmented RF models had errors of 10$\%$ and 16$\%$, respectively.

Overall, graph-based representations offer greater predictive accuracy and generalisability than conventional des-criptor-based approaches. Additionally, lattice perturbation shows promise as an augmentation strategy to enhance model performance in graph-based learning frameworks.  It should be noted that training CGCNN on SMOGN data, required matching the new descriptors with estimated compositions. Furthermore, the TC values generated by SMOGN are not experimentally verified. Furthermore, the lattice parameters used to generate the CIF were calculated using Equation \ref{eqtn_mouta_lattice_constant}, which does introduce a degree of error.

% Use the style of numbering in square brackets.
% If nothing is used, default style will be taken.
%\begin{enumerate}[a)]
%\item 
%\item 
%\item 
%\end{enumerate}  

%\begin{itemize}
%\item 
%\item 
%\item 
%\end{itemize}  

%\begin{description}
%\item[]
%\item[] 
%\item[] 
%\end{description}  

% Uncomment and use as the case may be
%\begin{theorem} 
%\end{theorem}

% Uncomment and use as the case may be
%\begin{lemma} 
%\end{lemma}


\printcredits

\section*{Declaration of competing interest}
\noindent The authors declare that they have no known competing financial interests or personal relationships that could have appeared to influence the work reported in this paper.

\section*{Acknowledgements}
\noindent This work was supported by the Engineering and Physical Sciences Research Council (EPSRC) (grant number EP/V010093/1).

\section*{Data availability}

The data and code used for this study are available on a public repository on Github (\href{https://github.com/Roms89/Data_representation_for_RE_oxide_thermal_conductivity}{link here}).

\bibliographystyle{cas-model2-names}

\bibliography{cas-refs}



\end{document}

\begin{frontmatter}

\title{A Methodological Study on Data Representation for Machine Learning Modelling of Thermal Conductivity of Rare-Earth Oxides}

\author[inst1]{Amiya Chowdhuryn}
\author[inst1]{Acacio Rinc\'on Romero}
\author[inst1]{Halar Memon}
\author[inst2]{Eduardo Aguilar-Bejarano}
\author[inst3]{Grazziela Figueredo}
\author[inst1]{Tanvir Hussain}

\address[inst1]{Centre for Excellence in Coatings and Surface Engineering, University of Nottingham, University Park, Nottingham, NG7 2RD, UK}

\address[inst2]{School of Chemistry, University of Nottingham, Jubilee Campus, Nottingham, NG8 1BB, UK}

\address[inst3]{Centre for Health Informatics, University of Nottingham, University Park, Nottingham, NG7 2RD, UK}

\beginsupplement{
    \setcounter{section}{0}
    \renewcommand{\thesection}{S\arabic{section}}
    \setcounter{equation}{0}
    \renewcommand{\theequation}{S\arabic{equation}}
    \setcounter{figure}{0}
    \renewcommand{\thefigure}{S\arabic{figure}}
    \setcounter{table}{0}
    \renewcommand{\thetable}{S\arabic{table}}

\newpage
\section*{Supplementary Material}\label{Supplementary}

Figure \ref{Img_theor_lattice_param} shows the theoretical and experimentally calculated (from XRD) lattice parameters for Lanthanum Gadolinium multi-component zirconates, where $x$ is the Gadolinium content of the composition. The equation is described in the main paper.

\begin{figure}[H]
    \centering
    \includegraphics[width=1\textwidth]{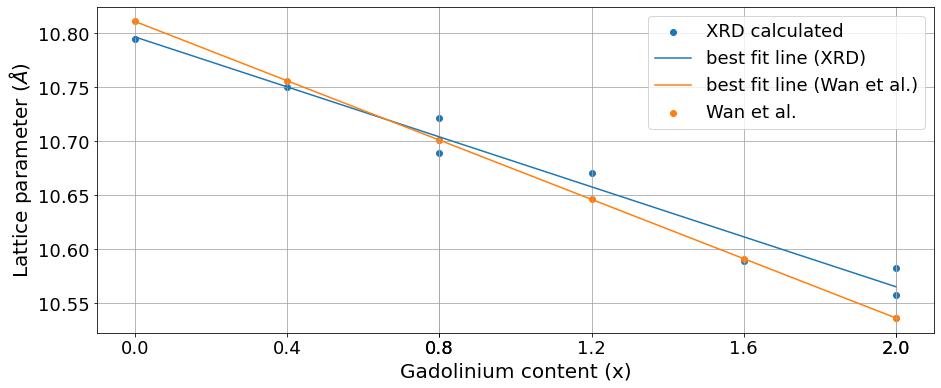}  
    \caption{Theoretically calculated lattice parameter vs experimentally calculated lattice parameter}
    \label{Img_theor_lattice_param}
\end{figure}

The correlation matrix below (Figure \ref{Img_Corr_matrix}), displays the correlation between each descriptor. Higher values indicate higher positive or negative correlation between two descriptors. $R_{A}$ 
 strong, positive correlation can be seen between Lanthanum and , while a strong, negative correlation can be seen between Ytterbium. $R_{B}$ is shown to have a strong positive and negative correlation with Titanium and Zirconia, respectively. Zirconia and Titanium have ionic radii of $0.72$ {\AA} and $0.605$ {\AA}, respectively. Hafnium is also present in the $B$ position and has an ionic radii of $0.71$ {\AA}. Despite this, the correlation factor is rather low with $R_{B}$. This can be explained by the distribution of the atomic fraction of Hafnium present in the data (figure \ref{Img_Wright_et_al_box_plot_comp}). Both Titanium and Zirconia are present in higher concentrations among the compositions present in the database, accounting for the higher correlation factors.
 
\begin{figure}[H]
    \centering
    \includegraphics[width=1\textwidth]{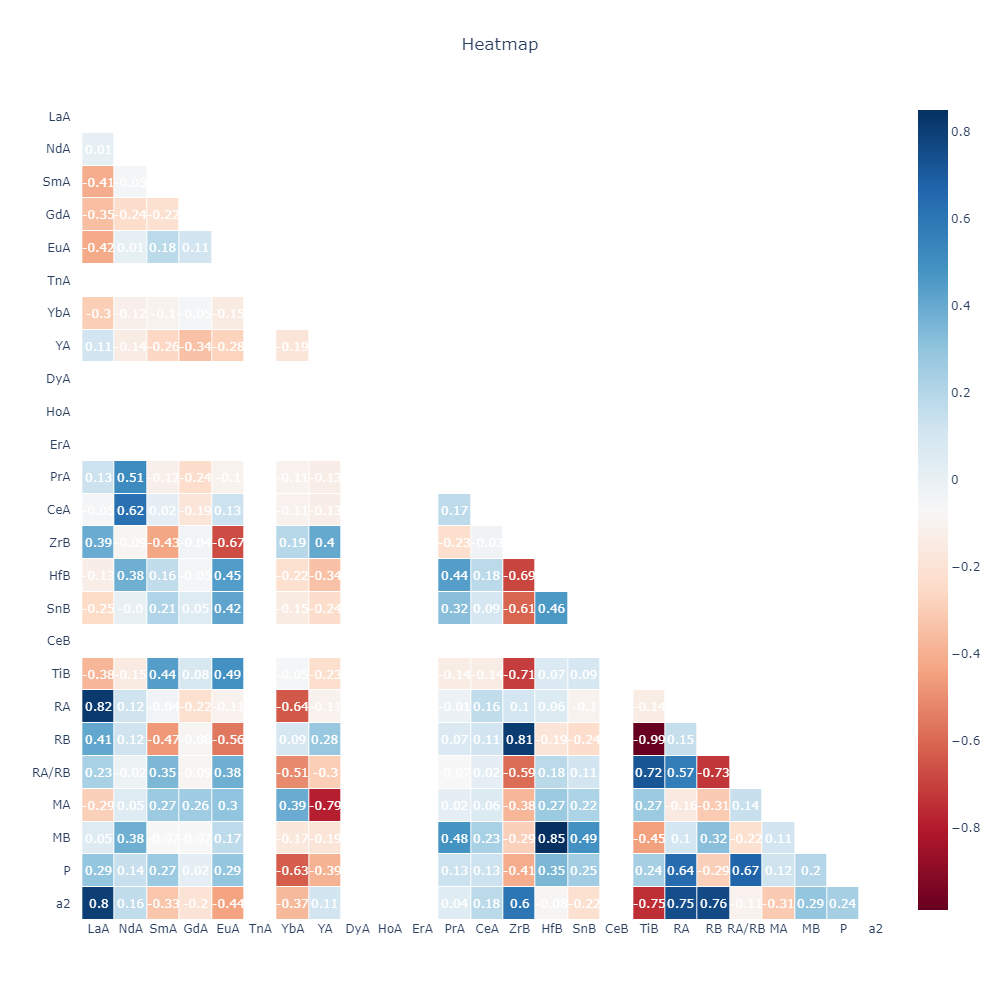}  
    \caption{Feature correlation matrix}
    \label{Img_Corr_matrix}
\end{figure}

Figures \ref{Img_Wright_et_al_box_plot_comp} and \ref{Img_TC_box_plot} show the distribution of values for the atomic fraction of each element, as well as the distribution of thermal conductivity values in the \textit{Wright et al.} database.

\begin{figure}[H]
    \centering
    \includegraphics[width=1\textwidth]{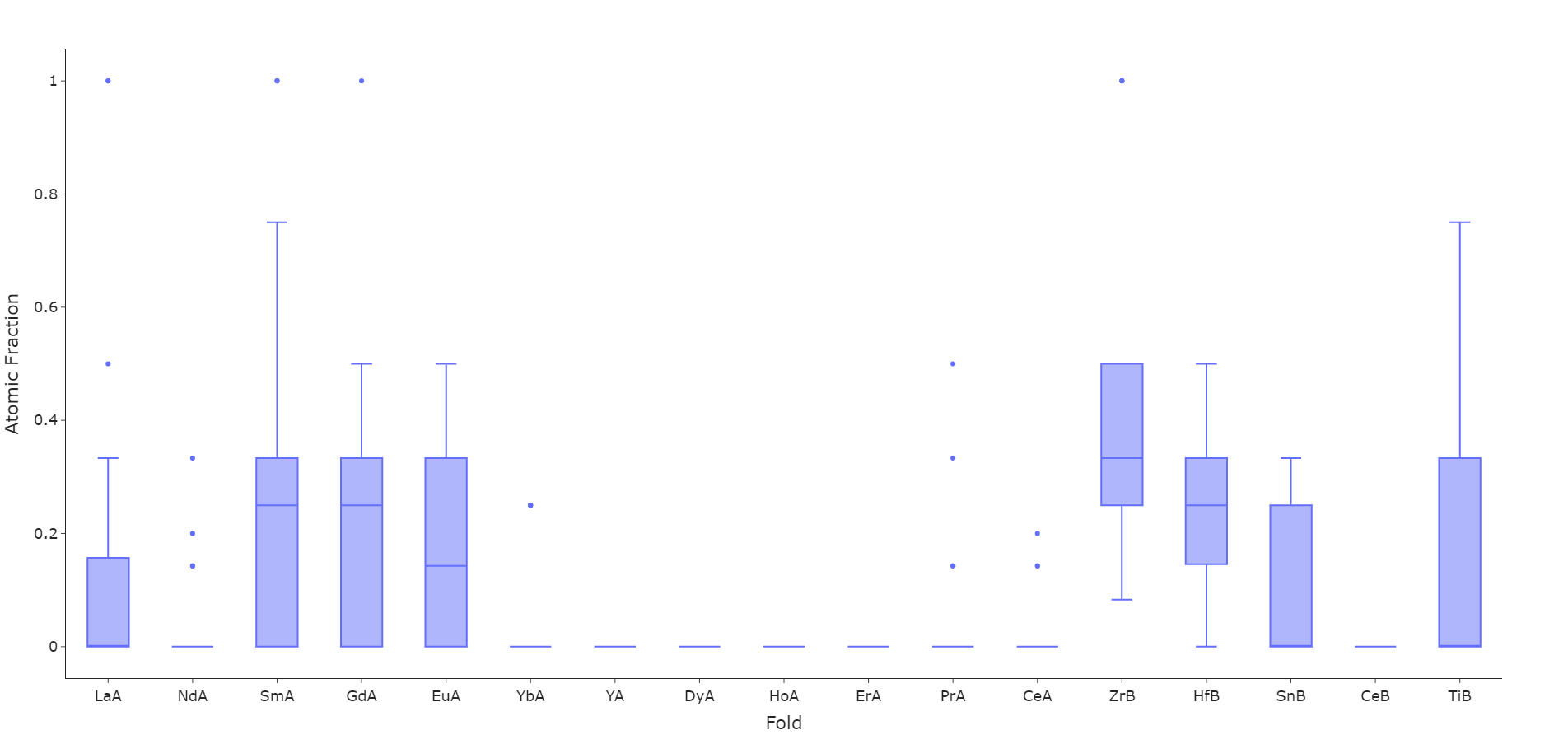}  
    \caption{Box plot showing distribution of values for each atomic species across all compositions in the Wright et al. database}
    \label{Img_Wright_et_al_box_plot_comp}
\end{figure}

\begin{figure}[H]
    \centering
    \includegraphics[width=1\textwidth]{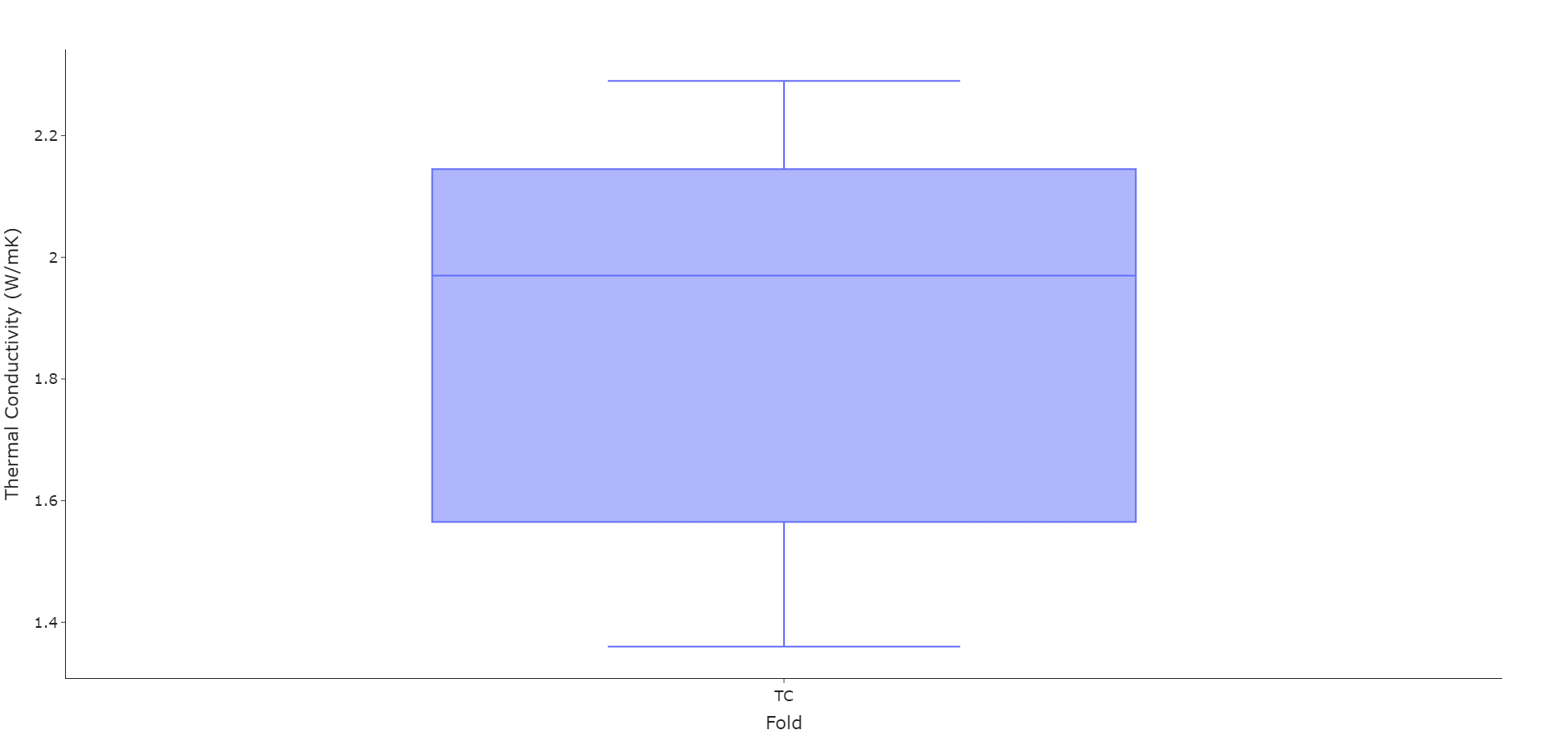}  
    \caption{Box plot showing distribution of values for thermal conductivity across all compositions in the Wright et al. database}
    \label{Img_TC_box_plot}
\end{figure}

Figure \ref{Img_SMOGN_data_distribution} compares the data distribution between the original data-set and the SMOGN augmented data-set.

\begin{figure}[H]
    \centering
    \includegraphics[width=1\textwidth]{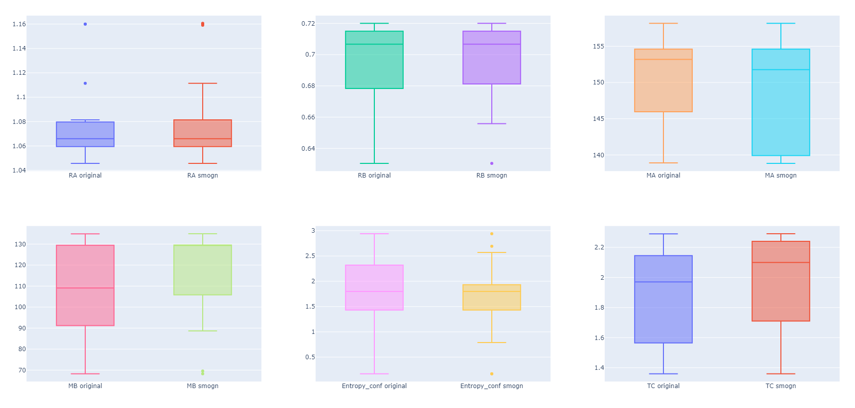}  
    \caption{Data distribution of database before (left) and after (right) adding the synthetically generated data points}
    \label{Img_SMOGN_data_distribution}
\end{figure}

Table \ref{Table_atom_features}, below, contains the atomic properties used by the GNN. The information is encoded in an array, using one-hot encoding, that is present in each node of the graph.

\begin{table}[H]
\begin{center}
    \begin{tabular}{ccc} 
         \hline
         Property & Unit & Range\\
         \cmidrule(lr){1-1}\cmidrule(lr){2-2}\cmidrule(lr){3-3}
         Group number & \_\ & {1, 2, ..., 18}\\
         Period number & \_\ & {1, 2, ..., 9}\\
         Electronegativity & \_\ & {0.5-4.0}\\
         Covalent radius & $pm$ & {25-250}\\
         Valence electrons & \_\ & {1, 2, ..., 12}\\
         First ionization energy & $eV$ & {1.3-3.3}\\
         Electron affinity & $eV$ & {$-$3-3.7}\\
         Block & \_\ & \textit{s, p, d, f}\\
         Atomic volume & $cm^{3}/mol$ & {1.5-4.3}\\
         \hline
    \end{tabular}
\caption{ Properties used in atom feature vector}
\label{Table_atom_features}
\end{center} 
\end{table}

\begin{landscape}
\begin{longtable}[c]{m{5cm}ccc}
\hline
\multicolumn{1}{c}{\textbf{Composition}} &
  \multicolumn{1}{c}{\textbf{Lattice Parameter}} &
  \multicolumn{1}{c}{\textbf{\begin{tabular}[c]{@{}l@{}}theoretical lattice density\\                                    (g/cm3)\end{tabular}}} &
  \multicolumn{1}{c}{\textbf{\begin{tabular}[c]{@{}l@{}}Measured Thermal \\
                                                        conductivity \\                     
                     ($W/(mK)$)\end{tabular}}} \\
  \cmidrule(lr){1-1}\cmidrule(lr){2-2}\cmidrule(lr){3-3}\cmidrule(lr){4-4}
\endfirsthead
\hline
\multicolumn{4}{c}%
{{\bfseries Table \thetable\ continued from previous page}} \\
\multicolumn{1}{c}{\textbf{Composition}} &
  \multicolumn{1}{c}{\textbf{Lattice Parameter}} &
  \multicolumn{1}{c}{\textbf{\begin{tabular}[c]{@{}l@{}}theoretical lattice density\\                                    (g/cm3)\end{tabular}}} &
  \multicolumn{1}{c}{\textbf{\begin{tabular}[c]{@{}l@{}}Measured Thermal \\
                        conductivity \\                     
                     ($W/(mK)$)\end{tabular}}} \\
  \cmidrule(lr){1-1}\cmidrule(lr){2-2}\cmidrule(lr){3-3}\cmidrule(lr){4-4}
\endhead
\rowcolor[HTML]{FFFFFF} 
La2Zr2O7                                                                & 10.728               & 6.135                & 2.140                \\
Sm 2 Zr 2   O 7                                                         & 10.574               & 6.665                & 2.040                \\
La 2 (Hf   1/2 Zr 1/2 ) 2 O 7                                           & 10.714               & 7.103                & 2.290                \\
Sm 2 (Sn   1/4 Ti 1/4 Hf 1/4 Zr 1/4 ) 2 O 7                             & 10.467               & 7.285                & 1.730                \\
Gd 2 (Sn   1/4 Ti 1/4 Hf 1/4 Zr 1/4 ) 2 O 7                             & 10.417               & 7.552                & 1.580                \\
(Sm 1/2 Gd   1/2 ) 2 (Ti 1/3 Hf 1/3 Zr 1/3 ) 2 O 7                      & 10.434               & 7.360                & 1.470                \\
(Eu 1/2 Gd   1/2 ) 2 (Ti 1/3 Hf 1/3 Zr 1/3 ) 2 O 7                      & 10.422               & 7.405                & 1.520                \\
(La 1/2 Pr   1/2 ) 2 (Sn 1/3 Hf 1/3 Zr 1/3 ) 2 O 7                      & 10.500               & 7.444                & 2.240                \\
(Eu 1/2 Gd   1/2 ) 2 (Sn 1/3 Hf 1/3 Zr 1/3 ) 2 O 7                      & 10.499               & 7.784                & 2.240                \\
(La 1/3 Pr   1/3 Nd 1/3 ) 2 (Hf 1/2 Zr 1/2 ) 2 O 7                      & 10.555               & 7.483                & 2.160                \\
(Sm 1/3 Eu   1/3 Gd 1/3 ) 2 (Hf 1/2 Zr 1/2 ) 2 O 7                      & 10.535               & 7.795                & 1.970                \\
(Sm 1/3 Eu   1/3 Gd 1/3 ) 2 (Sn 1/3 Hf 1/3 Zr 1/3 ) 2 O 7               & 10.512               & 7.723                & 2.210                \\
(Sm 1/3 Eu   1/3 Gd 1/3 ) 2 (Ti 1/4 Sn 1/4 Hf 1/4 Zr 1/4 ) 2 O 7        & 10.442               & 7.403                & 1.670                \\
(Sm 1/4 Eu   1/4 Gd 1/4 Yb 1/4 ) 2 (Ti 1/4 Sn 1/4 Hf 1/4 Zr 1/4 ) 2 O 7 & 10.403               & 7.604                & 1.450                \\
(Sm 1/3 Eu   1/3 Gd 1/3 ) 2 (Ti 1/2 Sn 1/6 Hf 1/6 Zr 1/6 ) 2 O 7        & 10.370               & 7.072                & 1.710                \\
(Sm 1/3 Eu   1/3 Gd 1/3 ) 2 (Ti 3/4 Sn 1/12 Hf 1/12 Zr 1/12 ) 2 O 7     & 10.297               & 6.728                & 1.900                \\
(Sm 1/3 Eu   1/3 Gd 1/3 ) 2 Ti 2 O 7                                    & 10.221               & 6.371                & 2.880                \\
(Sm 1/4 Eu   1/4 Gd 1/4 Yb 1/4 ) 2 (Ti 1/2 Hf 1/4 Zr 1/4 ) 2 O 7        & 10.344               & 7.310                & 1.360                \\
(Sm 3/4 Yb   1/4 ) 2 (Ti 1/2 Zr 1/2 ) 2 O 7                             & 10.370               & 6.686                & 1.400                \\
(La 1/5 Ce   1/5 Nd 1/5 Sm 1/5 Eu 1/5 ) 2 Zr 2 O 7                      & 10.637               & 6.431                & 2.060                \\
(La 1/7 Ce   1/7 Pr 1/7 Nd 1/7 Sm 1/7 Eu 1/7 Gd 1/7 ) 2 (Hf 1/2 Zr 1/2 ) 2 O 7 &
  10.565 &
  7.572 &
  1.970 \\
(La 1/7 Ce   1/7 Pr 1/7 Nd 1/7 Sm 1/7 Eu 1/7 Gd 1/7 ) 2 (Sn 1/3 Hf 1/3 Zr 1/3 ) 2 O 7 &
  10.543 &
  7.499 &
  2.020 \\
La1.6Y0.4Zr2O7                                                          & 10.676               & 6.007                & 1.760                \\
La1.2Y0.8Zr2O7                                                          & 10.622               & 5.877                & 1.902                \\
La0.8Y1.2Zr2O7                                                          & 10.567               & 5.745                & 2.307                \\
La0.4Y1.6Zr2O7                                                          & 10.510               & 5.609                & 2.379                \\
(La1/6Gd1/6Y2/3)2Zr2O7                                                  & 10.512               & 4.212                & 2.314                \\
(La1/6Gd2/3Y1/6)2Zr2O7                                                  & 10.547               & 5.178                & 2.156                \\
(La2/3Gd1/6Y1/6)2Zr2O7                                                  & 10.651               & 5.577                & 2.058                \\
(La1/3Gd1/3Y1/3)2Zr2O7                                                  & 10.571               & 4.997                & 2.244                \\
\hline
\multicolumn{1}{c}{}                                                    & \multicolumn{1}{c}{} & \multicolumn{1}{c}{} & \multicolumn{1}{l}{} \\
\multicolumn{1}{c}{} \\                                                   & \multicolumn{1}{c}{} & \multicolumn{1}{c}{} & \multicolumn{1}{l}{} \\
\caption{List of compositions used in training models}
\label{Table_composition_list}
\end{longtable}
\end{landscape}}